\documentclass[pre,twocolumn,twoside,showpacs,superscriptaddress]{revtex4-1}
\usepackage{graphics,amssymb,amsmath,epsf,color,hyperref}
\epsfclipon
\usepackage{epsfig,bm,epsf,graphics}
\usepackage{graphicx}
\usepackage{dcolumn}
\usepackage{bm}
\usepackage{color}

\begin{document}

\title{Data-driven inference of hidden nodes in networks}

\author{Danh-Tai Hoang}
\affiliation{Laboratory of Biological Modeling, National Institute of Diabetes and Digestive and Kidney Diseases, National Institutes of Health, Bethesda, Maryland 20892, USA}
\affiliation{Department of Natural Sciences, Quang Binh University, Dong Hoi, Quang Binh 510000, Vietnam}

\author{Junghyo Jo}
\email[Corresponding author: ]{jojunghyo@kmu.ac.kr}
\affiliation{School of Computational Sciences, Korea Institute for Advanced Study, Seoul 02455, Korea}
\affiliation{Department of Statistics, Keimyung University, Daegu 42601, Korea}

\author{Vipul Periwal}
\email[Corresponding author: ]{vipulp@mail.nih.gov}
\affiliation{Laboratory of Biological Modeling, National Institute of Diabetes and Digestive and Kidney Diseases, National Institutes of Health, Bethesda, Maryland 20892, USA}

\date{\today}

\begin{abstract}
The explosion of activity in finding interactions in complex systems is driven by availability of copious observations of complex natural systems. However, such systems, e.g. the human brain, are rarely completely observable. Interaction network inference must then contend with hidden variables affecting the behavior of the observed parts of the system. 
We present a novel data-driven approach for model inference with hidden variables. From configurations of observed variables, we identify the observed-to-observed, hidden-to-observed, observed-to-hidden, and hidden-to-hidden interactions, the configurations of hidden variables, and the number of hidden variables. We demonstrate the performance of our method by simulating a kinetic Ising model, and show that our method outperforms existing methods. Turning to real data, we infer the hidden nodes in a neuronal network in the salamander retina and a stock market network. We show that predictive modeling with hidden variables is significantly more accurate than that without hidden variables. Finally, an important hidden variable problem is to find the number of clusters in a dataset. We apply our method to classify MNIST handwritten digits. We find that there are about 60 clusters which are roughly equally distributed amongst the digits. 
\end{abstract}

 
 \maketitle


\section{Introduction}
To go from observations to predictive understanding is to go from stamp-collecting to science. Absent principled quantitative laws, biological and social systems can be generally described as networks of interacting nodes, with time-series data providing a window on the dynamics of the underlying system.
In the present era of big data, the network reconstruction problem has attracted considerable interest in research areas ranging from neuroscience~\cite{Schneidman2006,dombeck2007imaging,Nguyen2016, Bernal-Casas2017} and genomics~\cite{Lezon2006,Hickman2009,Bar2012} to finance ~\cite{Pincus2004, Tse2010,Tabak2010,Bury2012}. 
A fundamental caveat is that such reconstructions always rely on partial observation of these complex networks. For example, it is hopeless to follow the simultaneous spiking activity of every neuron in the brain, the transcription of every gene in the genome, and every fluctuating factor in a financial system. 

The problem of accounting for the unobserved constituents of any system is ill-posed without further information, simply because the number, the interactions, and the configurations of these hidden nodes must all be identified from the observed data and, a priori, one can make the former two as large and as complicated, respectively, as one pleases. To render the problem well-defined, one can first choose a theoretical model structure and then account for the unobserved nodes within this structure. Given the importance of this problem, much work has been devoted to it. 

A simple approach is to maximize the likelihood of observed configurations after marginalizing unobserved configurations~\cite{Roudi2013pre}. 
Another effective approach is the Expectation Maximization (EM) algorithm for hidden variables that contains two alternating steps, inferring all interactions of observed and hidden variables from configurations of observed variables, and reconstructing the configurations of hidden variables consistent with these inferred interactions~\cite{Dempster1977}. As one might expect, this algorithm is computationally impractical for even moderately large systems if the fraction of unobserved variables is significant. Furthermore, hidden variable configuration reconstruction accuracy is greatly dependent on interaction inference accuracy, a factor that becomes significant for limited datasets. Therefore, recent network reconstruction methods have considered alternative approaches, such as mean field approximations~\cite{Roudi2013pre, Hertz2014} and replica methods~\cite{Romano2014,Roudi2015jstat}.
However, the mean field approximations work only for weak and dense interactions~\cite{Hertz2014}, whereas the replica methods  allows to infer strong and sparse interactions, but impose the stringent assumption of the independence between hidden variables~\cite{Roudi2015jstat}.
In addition to non-interacting hidden variables, random interaction strengths and the thermodynamic limit are two prerequisites for the exact inference of the replica methods~\cite{Romano2014}.

We recently formulated a new approach~\cite{Hoang2018,Hoang2018codefem} to network reconstruction for observed variables that is significantly more accurate inference-wise in the limit of sparse sampling and orders of magnitude faster computation-wise than previous methods. Based on this foundation, we propose a new approach for network reconstruction including hidden variables, by replacing the inference step with our approach. This does not, by itself, address the crucial question of the number of unobserved variables, so we complete our proposal by formulating a simple quantitative test of model complexity to determine this number.

This paper is organized as follows: We briefly review our inference method and outline its extension to hidden variables, paying especial attention to the determination of the number and interactions of hidden variables, amongst themselves and with observed variables. We then validate our method with simulated data from kinetic Ising models, showing the accurate determination of the number and interactions of hidden variables for a range of observed fractions of systems, going up to $40\%$ hidden variables. Turning to real data, we apply our method to reconstruct a neural network from partially observed neuronal activities, and a stock-market network using data of opening and closing stock prices of 25 American companies. We validate our network reconstructions by reproducing observed neuronal activities by pinning just a few neuron configurations, and by exhibiting a profitable stock trading strategy based on our inferred network. Finally, we demonstrate that our approach is suited to unsupervised data clustering, as well, since cluster membership is a type of hidden variable. We estimate the number of hidden features that can explain the MNIST hand-written digit dataset. Complete source code with documentation is available~\cite{Hoang2018codehidden}.


 \section{Method}
 \label{method}
 
We explain our approach in the context of a concrete example for ease of understanding. Consider a stochastic dynamical system in which a vector of $N$ binary ($\pm 1$) variables $\bm{\sigma} = (\sigma_1, \cdots, \sigma_N)$ evolves stochastically according to the conditional probability:
\begin{equation}
\label{eq:kIsingProb}
P(\sigma_i(t+1)|\bm{\sigma}(t)) = \frac{\exp(\sigma_i(t+1) H_i(t))}{\exp(H_i(t)) + \exp(-H_i(t))},
\end{equation}
for $i = 1, \cdots, N$.
The local field $H_i(t)=\sum_j W_{ij} \sigma_j(t)$ represents the summed influence of the present state $\sigma_j(t)$ on the future state $\sigma_i(t+1)$ through the weight $W_{ij}.$
This kinetic Ising model has a model expectation, $\langle  \sigma_i(t+1) \rangle_{\text{model}} = \tanh H_{i}(t)$.
Generating $\bm{\sigma}(t)$ given $W_{ij}$ is easy, but inferring $W_{ij}$ given $\bm{\sigma}(t)$ is not trivial.
Although numerous methods exist for the inverse problem~\cite{Roudi2011,Mezard2011,Zeng2013}, we recently proposed a new approach~\cite{Hoang2018,Hoang2018codefem}. 
We give here a simplified intuitive account. The first step is the linear regression of $H_i = \sum_j W_{ij} \sigma_j$ between $H_i$ and $\sigma_j$.
Suppose we know $H_i(t)$ and $\sigma_j(t).$ The coefficient $W_{ij}$ can then be obtained as usual:
\begin{equation}
\label{eq:w}
W_{ij} = \sum_k \langle \delta H_i \delta \sigma_k \rangle [C^{-1}]_{kj},
\end{equation}
where $C_{jk} \equiv \langle \delta\sigma_j\delta\sigma_k\rangle$ is the covariance matrix for $\bm{\sigma}(t),$
 with $\langle f \rangle \equiv L^{-1} \sum_{t=1}^L f(t)$ and $\delta f \equiv f - \langle f \rangle.$ 
The second step is the update of the observable,
\begin{equation}
\label{eq:H}
H_i(t) \leftarrow  \frac{\sigma_i(t+1)}{\langle \sigma_i(t+1) \rangle_{\text{model}}} H_i(t) = \sigma_i(t+1) \frac{H_i(t)}{\tanh H_i(t)}.
\end{equation}
The multiplicative update of $H_i(t)$ corrects the magnitude and sign of $H_i(t)$ based on the ratio of observed $\sigma_i(t+1)$ and model expectation $\langle \sigma_i(t+1) \rangle_{\text{model}},$ which is always larger than unity in absolute magnitude. A critical aspect of Eq.~(\ref{eq:H}) is that the limit $|H_i|\downarrow 0$ gives $H_i(t) \leftarrow  \sigma_i(t+1),$ independent of $H_i.$ Therefore, the update in Eq.~(\ref{eq:H}) avoids being entirely multiplicative for determining $W_{ij}.$
These two steps, $H_i(t) \rightarrow W_{ij}$ and $W_{ij} \rightarrow H_i(t),$ provide a powerful iterative method.
We continue this iteration until the discrepancy between data and model expectation $D_i(W)\equiv\sum_{t} \big[ \sigma_i(t+1) - \langle \sigma_i(t+1) \rangle_{\text{model}} \big]^2$ is minimized.
We derived the linear regression in Eq.~(\ref{eq:w}) using the concept of free energy in statistical mechanics~\cite{Hoang2018} so we call this method Free Energy Minimization (FEM). Notice that the parameter update in Eqs.~(\ref{eq:w}-\ref{eq:H}) is completely independent of the computation of $D_i.$ This crucial feature allows the small sample size inference to avoid overfitting because the minimization of $D_i$ is used only as a stopping criterion.

Now we propose to apply the FEM method to infer interactions from/to hidden variables.
The system has $N_v$ observable (visible) and $N_h$ hidden variables ($N=N_v+N_h$).
As a variant of the EM algorithm, we first assign random configurations for hidden variables.
We then infer interaction weights $W_{ij}$ for observed-to-observed, hidden-to-observed, observed-to-hidden, and hidden-to-hidden variables with the FEM method.
Given $W_{ij}$, we can update the configurations of hidden variables with a probability $\mathcal{L}_{2} /(\mathcal{L}_{1}+\mathcal{L}_{2})$ where $\mathcal{L}_{1}$ and $\mathcal{L}_{2}$ represent the likelihoods $\mathcal{L}$ of the system before and after flipping,
\begin{equation}
\label{eq:L}
{\cal{L}} = \prod_{t=1}^{L-1}\prod_{i=1}^{N} P(\sigma_i(t+1)|\bm{\sigma}(t)).
\end{equation}
Note that the independent terms in the update of hidden states (at each $t$) of $\mathcal{L}_{1}$ and $\mathcal{L}_{2}$ cancel in the update ratio, $\mathcal{L}_{2} /(\mathcal{L}_{1}+\mathcal{L}_{2}).$ Therefore, we just need to calculate the dependent terms. 
The iterations between the parameter optimization (M step) and the variable update (E step) provide accurate inference of the interaction weights, $W_{ij},$ and the unknown configurations of hidden variables.

We must now consider the problem of determining the number of hidden variables. 
A simple measure would be the same discrepancy between observation $\sigma^v_i(t+1)$ and model expectation $\langle \sigma^v_i(t+1) \rangle_{\text{model}},$
\begin{equation}
\label{eq:Dobs}
D_v \equiv \sum_{i=1}^{N_v} D_i(W) 
\end{equation}
but this is clearly not taking the hidden variables into account. On the other hand, extending the sum in Eq.~(\ref{eq:Dobs}) to include hidden variables is useless because the E step update is minimizing these additional terms already. 
Since the error in inference of hidden variable states cannot be set by a scale smaller than the model discrepancy in the observed part, we define the scaled discrepancy of the entire system based on the observed part as 
\begin{equation}
\label{eq:D}
D \equiv D_v \bigg(1+\frac{N_h}{N_v}\bigg).
\end{equation}
The first term in Eq.~\ref{eq:D} represents the goodness of fit for observed variables and the second term represents model complexity, so our criterion balances the two. Because $D_v \varpropto - \log \mathcal{L}_v$ where $\mathcal{L}_v$ represents the likelihood of observed variables, Eq.~\ref{eq:D} can be rewritten as  
$D \varpropto - \log \mathcal{L}_v (1+N_h/N_v)$. Our criterion is thus similar in spirit to the Akaike information criterion~\cite{Akaike1974} and Bayesian information criterion~\cite{Schwarz1978} with log-likelihood of observation ($- \log \mathcal{L}_v \sim D_v$) and model degrees of freedom ($N_v+N_h$).

Finally, our method can be summarized as the following set of steps:\\
For a range of numbers of hidden variables, in parallel and independently,\\
(i) Assign configurations of hidden variables at random; \\
(ii) Infer interaction weights $W_{ij}$ including observed-to-observed, hidden-to-observed, observed-to-hidden, and hidden-to-hidden from the configurations of observed and hidden variables using FEM; \\
(iii) Flip the states of hidden variables with probability $\mathcal{L}_{2} /(\mathcal{L}_{1}+\mathcal{L}_{2})$ (see Eq.~(\ref{eq:L})). \\
(iv) Repeat steps (ii) and (iii) until the discrepancy of observed variables is minimized. The final values of $W_{ij}$ and hidden states are the inferred coupling weights and configurations of hidden spins, respectively.\\
Pick the number of hidden variables that minimizes Eq.~(\ref{eq:D}).

\section{Results}
\subsection{Kinetic Ising model}

\begin{figure*}
\centering
\includegraphics[width=16cm]{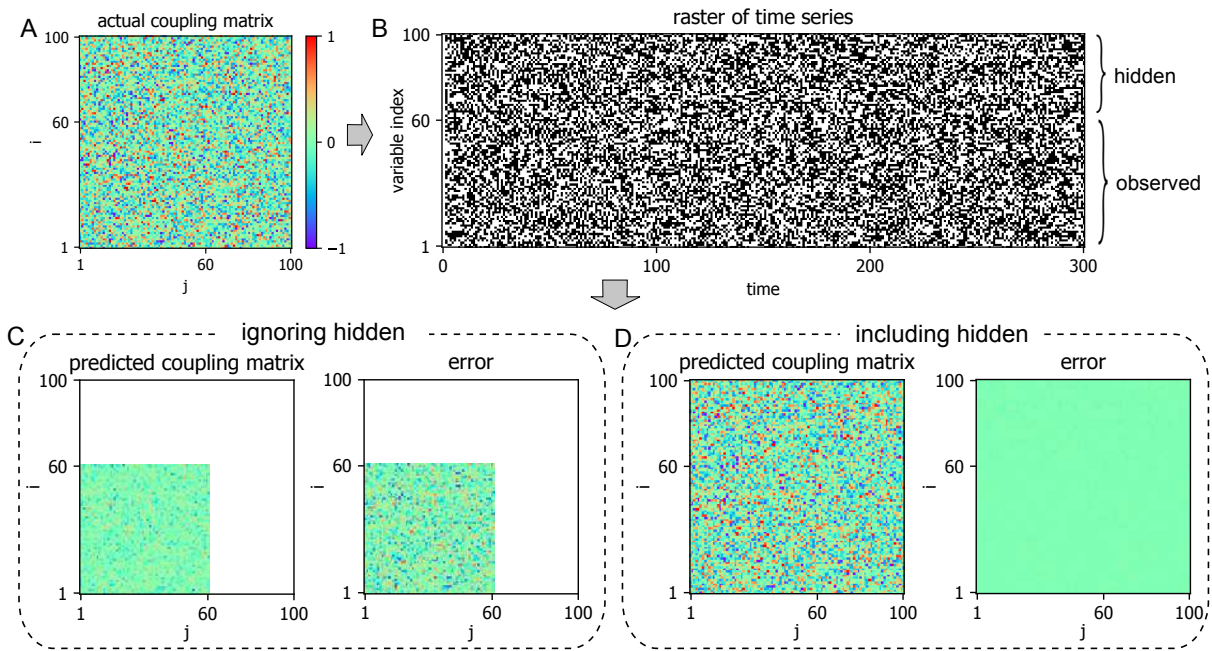}
\caption{ \label{fig1} (Color online) Network reconstruction from partial observations. From the actual interaction weights (A), typical time series of 100 variables are generated according to the kinetic Ising model (B). Using the configuration of 60 observable variables, the interaction weights are recovered in two cases: ignoring (C) and including (D) the existence of hidden variables. Data length $L=40,000$ is used.
}
\end{figure*}

\begin{figure}
\centering
\includegraphics[width=5cm]{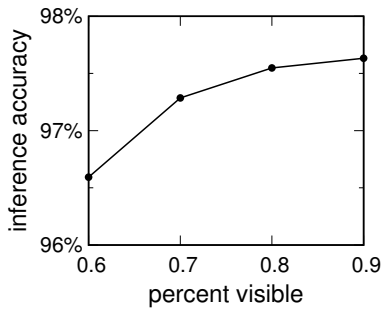}
\caption{ \label{fig_accuracy} Accuracy of network reconstruction and fraction of hidden variables.
The inference accuracy is plotted as a function of the fraction of visible variables, $N_v/N$.
System size $N = 100$ and data length $L=40,000$ are used. 
}
\end{figure}

To demonstrate the performance of our method, we synthesized binary time series of $N=100$ spins by using the Sherington-Kirkpatrick model~\cite{Sherrington1975}. The update of spin $\bm{\sigma}$ follows 
Eq.~(\ref{eq:kIsingProb}) with preset coupling strengths $W_{ij}$ (Fig.~\ref{fig1}A). Our goal is to reconstruct all of $W_{ij}$ from observations of a fraction of $\bm{\sigma}(t).$ 
Suppose that we only observe the time series of 60 spins with 40 spins hidden (Fig.~\ref{fig1}B).
When we reconstructed the interactions $W_{ij}$ between observed node $i$ and observed node $j$  using FEM, the reconstructed $W_{ij}$, based on the partial observations, showed a large error (Fig.~\ref{fig1}C). 
We introduced 40 hidden variables, and applied the EM algorithm outlined in the previous section.
The reconstructed $W_{ij}$ was  close to the true $W_{ij}$ (Fig.~\ref{fig1}D). 
How well are the hidden variable configurations recovered? 
For the case of 40 hidden variables, the true configurations of the hidden variables were recovered with an accuracy of 96.6\%.
The reconstruction accuracy increased with fewer spins hidden (Fig.~\ref{fig_accuracy}).
For instance, when 90 spins were observed with 10 spins hidden, the accuracy was 97.6\%.


\begin{figure*}
\centering
\includegraphics[width=16cm]{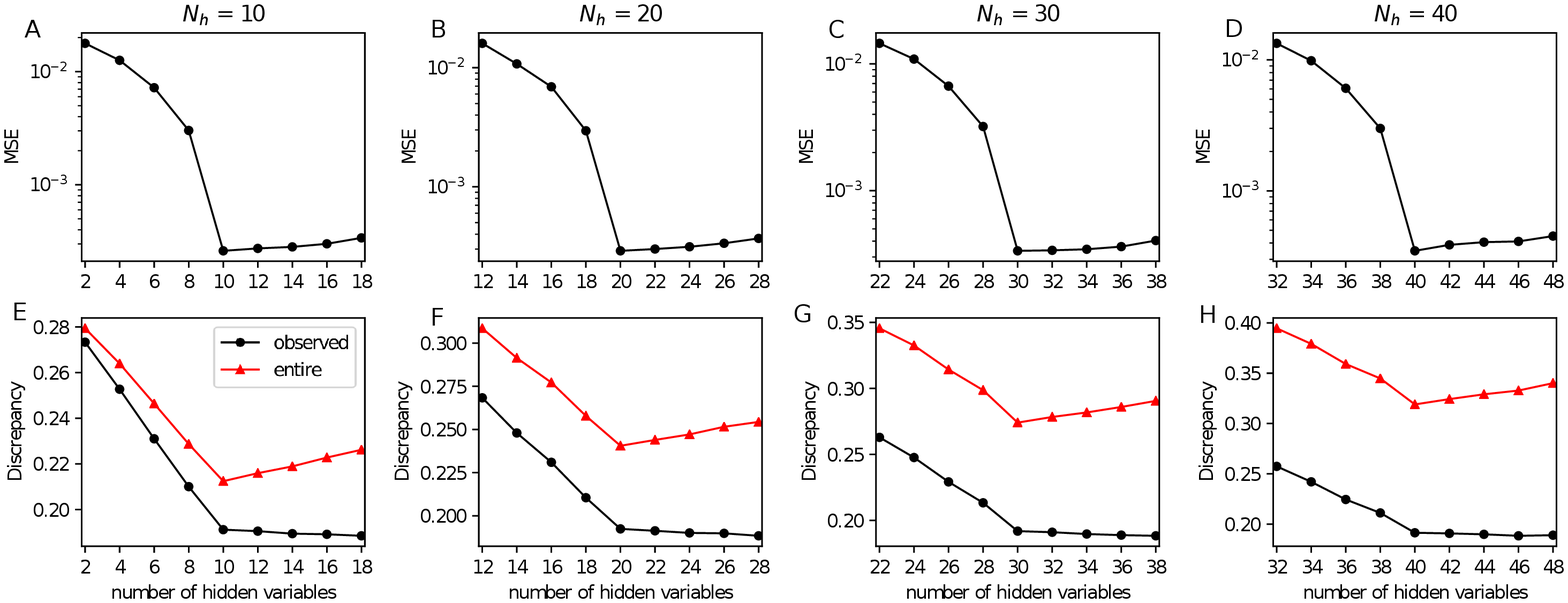}
\caption{ \label{fig_nh} (Color online) Estimation of hidden degrees of freedom. Mean square errors of observed variables (upper) and discrepancy $D_v$ of observed variables (lower, black circles) and discrepancy $D$ of total (observed and hidden) variables (lower, red triangles) are shown with differently assumed numbers $N_h$ of hidden variables. The actual numbers of hidden variables are $N_h=10$, $20$, $30$ and $40$, from left to right. A system size $N=100$ and data length $L=40,000$ are used.
}
\end{figure*}

The number of hidden variables is usually unknown in real-world problems.
When we reconstructed $W_{ij}$ with different numbers of hidden variables, the mean square errors of observed-to-observed interaction strengths, MSE$ = N_v^{-2} \sum_{i,j \in \text{obs}} (W_{ij} - W_{ij}^{\textrm{true}})^2$,
were minimal at the right number of hidden variables (Fig.~\ref{fig_nh}, upper panel).
The MSE is also inaccessible in real-world problems, but 
the minimum of $D$ (Eq.~(\ref{eq:D}))  captured the correct value of $N_h$ (Fig.~\ref{fig_nh}, lower panel, red lines).

\begin{figure*}
\centering
\includegraphics[width=16cm]{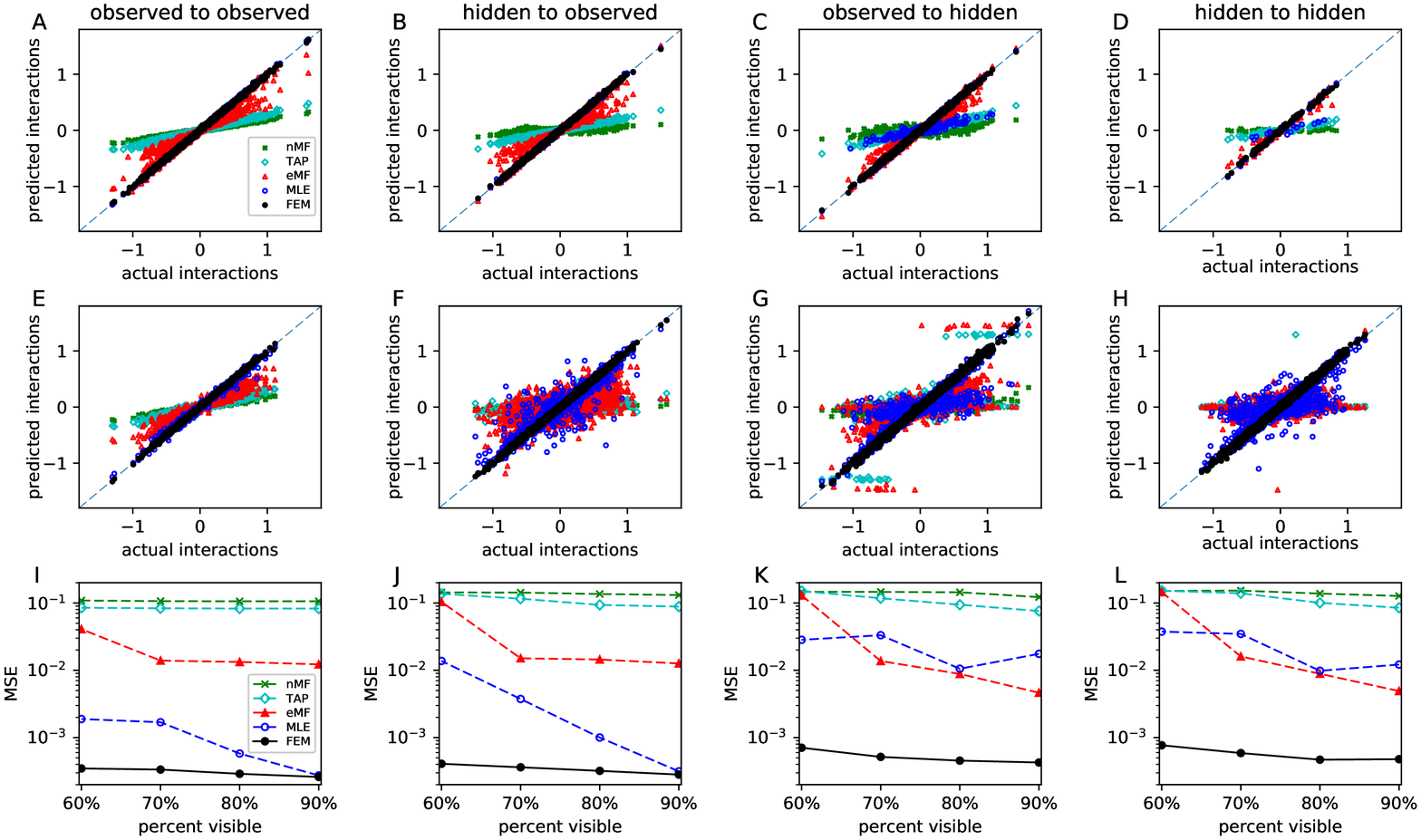}
\caption{ \label{fig_compare} (Color online) Performance comparison between inference methods. Predicted interactions versus actual interactions, separately in observed-to-observed (A and E), hidden-to-observed (B and F), observed-to-hidden (C and G), and hidden-to-hidden (D and H), for two numbers of hidden variables $N_h=10$ (first row) and $40$ (second row). The mean square errors between predicted interactions and actual interactions are shown as a function of the fraction ($N_v/N$) of observed variables over total variables (I-L). We compared five inference methods: na\"ive mean-field (nMF), Thouless-Anderson-Palmer (TAP), exact mean-field (eMF), Maximum Likelihood Estimation (MLE), and Free Energy Minimization (FEM). System size $N = 100$ and data length $L=40,000$ are used. For MLE, we used a learning rate $\alpha=1$.
}
\end{figure*}

To reconstruct $W_{ij}$ from observed and hidden variables, we used FEM.
For the M step, mean field methods such as na\"ive, Thouless-Anderson-Palmer, and exact mean field methods (nMF, TAP, and eMF), and maximum likelihood estimation (MLE) can also be used. A brief review of these methods can be found in Ref.~\cite{Hoang2018}. 
Given partial observations,  mean field approaches were not successful in reconstructing $W_{ij}$
(Fig.~\ref{fig_compare}).
For a small percentage of hidden variables (90 observable and 10 hidden), FEM and MLE showed a similar performance in the reconstruction of observed-to-observed and hidden-to-observed interactions.
However, FEM outperformed MLE in reconstructing observed-to-hidden and hidden-to-hidden interactions (Fig.~\ref{fig_compare}A-D).
For a large percentage of hidden variables (60 observable and 40 hidden variables), FEM showed significantly better performance even for observed-to-observed and hidden-to-observed interactions (Fig.~\ref{fig_compare}E-H).
We quantified the reconstruction performance by measuring MSE between $W_{ij}$ and $W_{ij}^{\text{true}}$.
FEM showed more accurate reconstruction of $W_{ij}$ with lower MSE in every case.
More importantly, in addition to better performance, FEM took approximately 100 times less computation time than MLE due to its multiplicative update (see~\cite{Hoang2018,Hoang2018codefem,Hoang2018codehidden} for details).


\subsection{Neuronal network}
\label{section_neuron}

\begin{figure*}
\centering
\includegraphics[width=16cm]{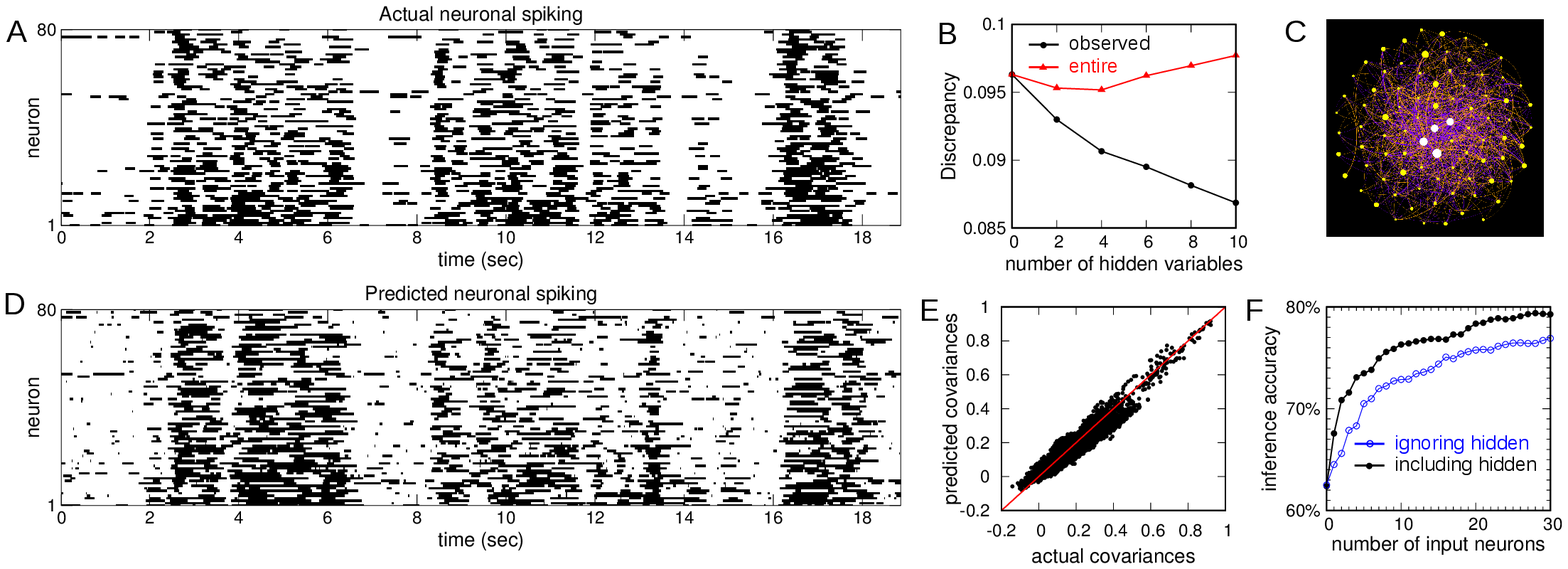}
\caption{ \label{fig_neuron} (Color online) Neural network reconstruction with hidden nodes. Activities of the 80 most active neurons are plotted with black dots representing active states and white dots representing silent states (A). Discrepancies between observed and expected neuronal activities (black circles) and discrepancies between entire neuronal activities and their expectations (red triangles) are shown as a functions of the number of hidden variables (B). A predicted neuronal network is visualized in which green nodes represent observed neurons, while white nodes represent hidden variables. The red and blue edges represent positive and negative couplings, respectively. Edge direction is clock-wise. The node size scales with value of firing rate, and edge thickness scales with coupling weight (C).
Given the reconstructed coupling strengths, external local fields and configurations of hidden variables, the activities of 80 neurons are reconstructed (D). Inferred covariances $C_{ij}$ versus actual covariances $C_{ij}^{\text{true}}$ (E). Inference accuracy of remaining neuronal activities are shown as a function of the number of input neurons for two cases: ignoring (blue) and including (black) the existence of hidden variables (F).
}
\end{figure*}

The analysis of real data brings out issues far more clearly than simulated validations. Therefore, we applied our method to infer hidden nodes and their contributions in a real neuronal network.
We used the time series data of the 80 most active neurons from published multi-channel recordings of neuronal firing in the salamander retina~\cite{Tkacik2014}. 
Considering the existence of unobserved hidden neurons, we modeled the evolution of neuronal activities by defining a local field, $H_{i}(t) =H_{i}^{\text{ext}} + \sum_{j} W_{ij}\sigma_{j} (t)$, that determines the future activity of $\sigma_i(t+1)$.
The external local field $H_{i}^{\text{ext}}$ represents the bias of the $i$th neuron that sets its threshold.
For various numbers of hidden neurons $N_h$, we computed $H_{i}^{\text{ext}}$ and $W_{ij}$.
The activities of observed neurons were explained better and $D_v$ kept decreasing with a larger $N_h$ of hidden neurons (Fig.~\ref{fig_neuron}B). However, once we considered the overall discrepancy $D,$ an optimal number of hidden neurons was $N_h^* = 4.$
Thus, the inclusion of four hidden neurons best explained the activities of observed neurons, taking model complexity into account. Given these four hidden neurons, the connection weights $W_{ij}$ were reconstructed as shown in Fig.~\ref{fig_neuron}C. 

Since $W_{ij}^{\text{true}}$ is unknown for the neuronal network, we validated our reconstruction in two different ways.
First, we selected some neurons as input neurons, and then based on the activities of these input neurons, we generated the activities of remaining neurons by using the reconstructed $W_{ij}$. 
Here we selected the input neurons based on having the strongest influence to other neurons by gauging $\sum_i |W_{ij}|$. Given varying numbers of input neurons, we could successfully reconstruct the actual activities of the remaining neurons (Fig.~\ref{fig_neuron}D). 
As the number of input neurons increased, the reconstruction accuracy increased (Fig.~\ref{fig_neuron}F). 
Moreover, once the four hidden neurons were considered the reconstruction accuracy was significantly improved.
Second, given $\bm{\sigma}(t)$, we predicted $\bm{\sigma}(t+1)$, and then calculated the covariance $C_{ij} = \langle \delta \sigma_i(t+1) \delta \sigma_j(t) \rangle $. 
The reconstructed covariance was comparable with the actual covariance from the observation (Fig.~\ref{fig_neuron}E). 

\subsection{Stock network}
\label{section_stock}
\begin{figure*}
\centering
\includegraphics[width=16cm]{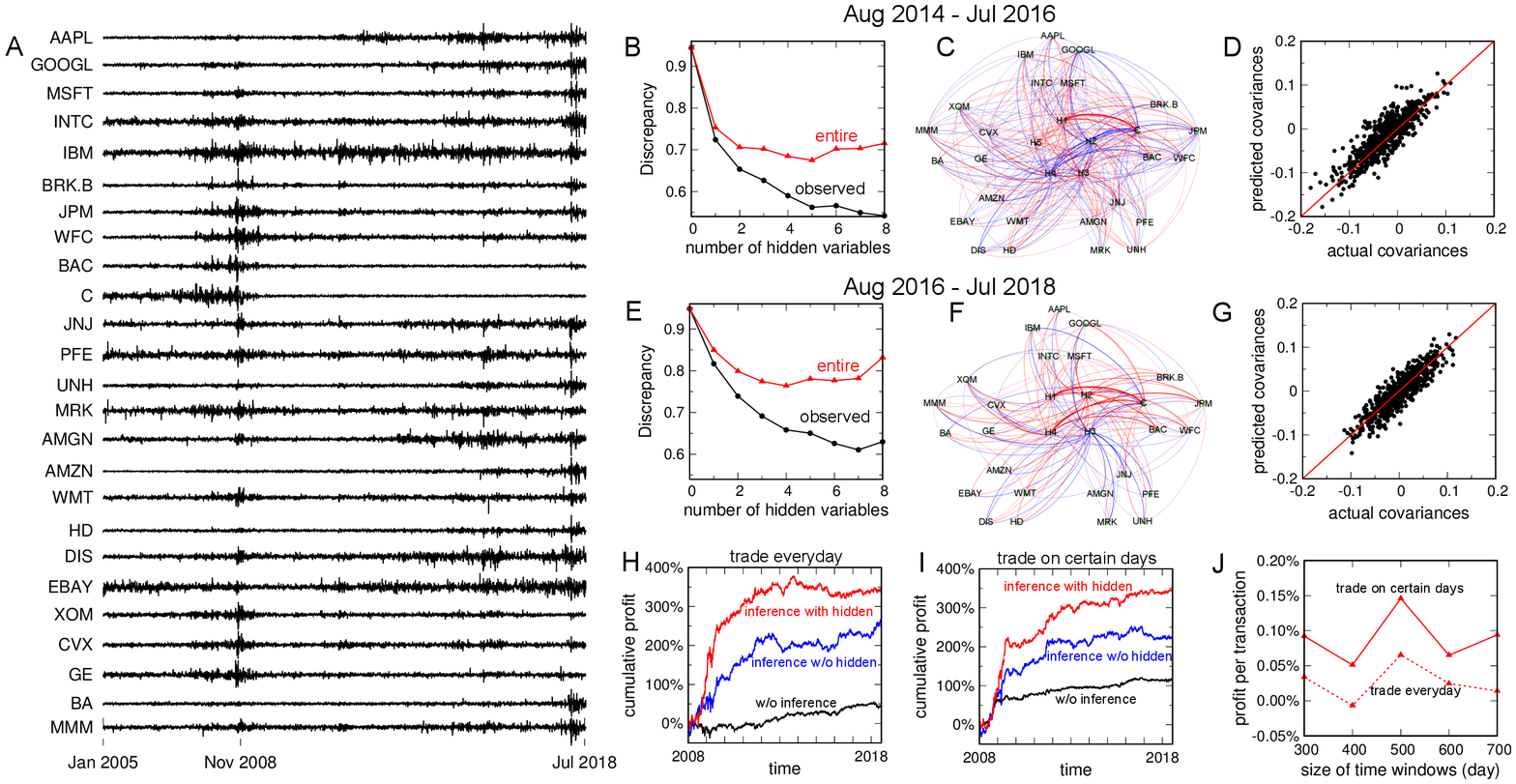}
\caption{ \label{fig_stock} (Color online) Stock-market network reconstruction with hidden nodes. The time series of the difference between opening and closing prices of 25 American companies are shown from January 2005 to July 2018 (A). The network reconstruction considered for different periods: from August 2014 to July 2016 (B, C, D), and from August 2016 to July 2018 (E, F, G). Discrepancies between observed and expected configurations (black circles) and discrepancies between entire variables and their expectations (red triangles) are computed to estimate the necessary number of hidden variables (B, E). Reconstructed stock-market networks are visualized (C, F) in which the red and blue edges represent positive and negative couplings, respectively. Edge direction is clock-wise, and edge thickness scales with coupling strengths. Inferred covariances $C_{ij}$ versus actual covariances $C_{ij}^{\text{true}}$ (D, G). Cumulative profits are shown as a function of the time period for trade on everyday (fully trade) (H) and on certain days (alternative trade) (I) with different trade strategies: random trades (black), strategic trades ignoring (blue) and including hidden variables (red). Profit per transaction versus time window size for the network reconstruction (J), for everyday trade (dashed line) and certain-day trade (solid line).
}
\end{figure*}

Our method has a wide range of practical applications.
As a demonstration, we reconstruct a stock market network with possible hidden nodes.
We used stock price time series of 25 major companies in the S\&P 500 index in  five different sectors: technology (AAPL, GOOGL, MSFT, INTC, IBM), finance (BRK.B, JPM, WFC, BAC, C), health care (JNJ, PFE, UNH, MRK, AMGN),  consumer discretionary (AMZN, WMT, HD, DIS, EBAY), and energy and industrial (XOM, CVX, GE, BA, MMM)~\cite{stockdata}. 
We examined the price difference between daily opening and closing stock prices. Their fluctuations from January 2005 to July 2018 are shown in (Fig.~\ref{fig_stock}A).
First, instead of considering the continuous price fluctuations, we defined a discretized measure of price changes.
If the daily price increased at time $t$ for the $i$th company (opening price $<$ closing price), we defined $\sigma_i(t) = +1$. However, if the price decreased (opening price $>$ closing price), then $\sigma_i(t) = -1$. Finally, if the price was unchanged (opening price $=$ closing price), we defined $\sigma_i(t) = \sigma_i(t-1)$.

We applied our method to this discretized data, and inferred external factors $H_{i}^{\text{ext}}$ and interacting factors $\sum_{j} W_{ij}\sigma_{j}(t)$ that stochastically determine $\sigma_i(t+1)$.
Since FEM works well even for small sample sizes~\cite{Hoang2018}, we divided the data into two-year periods to probe possible slower temporal changes in the interactions $W_{ij}$ between stock prices.
In particular, we show results from more recent data for 2014 to 2016 (Fig.~\ref{fig_stock}B-D) and 2016 to 2018 (Fig.~\ref{fig_stock}E-F). The discrepancy $D_v$ between observed $\sigma^v_i(t+1)$ and model expectation $\langle \sigma^v_i(t+1) \rangle_{\text{model}}$ kept decreasing as expected when more hidden nodes were introduced (Fig.~\ref{fig_stock}B and E). 
However, the entire discrepancy $D$, considering the model complexity with hidden variables, showed a minimum at $N^*_h \approx 4$-$5$ hidden nodes.
The inferred stock market network including interactions between observed and hidden nodes is visualized in Fig.~\ref{fig_stock}C and F.
When we generated time series of stock prices using the reconstructed network, we found that the covariance of the generated sequences was consistent with the covariance of original sequences (Fig.~\ref{fig_stock}D and G).


An accurate predictive network reconstruction should enable profitable trades. In particular, does our discrete reduction of the price data still contain enough information to be useful?
First, we reconstructed the interactions between companies including the appropriate number of hidden nodes by using stock price data for the most recent $T$ days: $\bm{\sigma}^v(t-T+1), \bm{\sigma}^v(t-T+2), \cdots, \bm{\sigma}^v(t)$.  Then, we predicted the price change direction $\bm{\sigma}^v(t+1)$ for the next day.
Our strategy was to buy the stock $i$ that had the highest probability of increasing with a maximum $H_i(t)$, 
and to sell the stock $j$ that had the highest probability of decreasing with a minimum $H_j(t)$ at the beginning of the day.
This trading strategy is expected to have a maximum profit bounded by (close price($i$) $-$ open price($i$)) $+$ (open price($j$)  $-$ close price($j$)).
The trading simulation from 2008 to 2018 with a moving time window $T=500$ days obtained 350\% cumulative profit (Fig.~\ref{fig_stock}H). This profit was significantly higher than the profit of 50\%  using random trades, which is due to the secular rise of the entire stock index. 
Furthermore, the reconstructed network including hidden nodes showed a larger profit than the 250\% profit from the reconstructed network ignoring the hidden nodes.
Next, we refined the trading strategy by buying/selling the stock that has the highest probability of increasing/decreasing but only if its price has decreased/increased on the previous day. In particular, this may result in only buying or only selling on any specific day. 
This new strategy produced the same cumulative profit in total, but it doubled the profit per transaction (Fig.~\ref{fig_stock}I and J). 
Finally, we confirmed that the optimal time window for the highest profit was about $T=500$ days (Fig.~\ref{fig_stock}J).

\subsection{Classification of handwritten digits}
\label{section_mnist}

\begin{figure*}
\centering
\includegraphics[width=16cm]{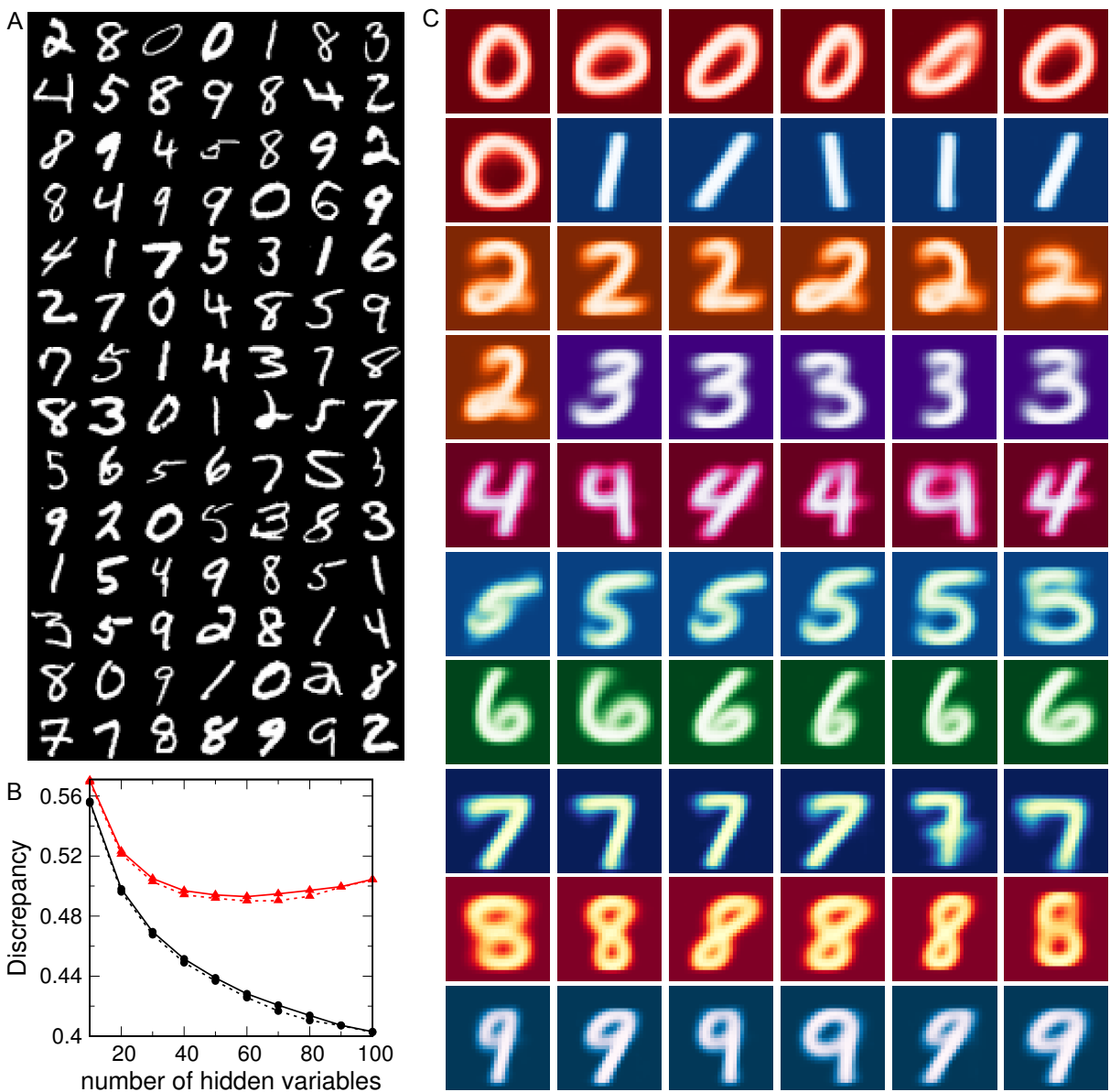}
\caption{ \label{fig_mnist} 
(Color online) Hidden degrees of freedom for the classification of handwritten digit images. (A) 98 image samples were randomly selected from the training set of the MNIST data. (B) Discrepancies between observed and expected configurations (black circles) and discrepancies between entire variables and their expectations (red triangles) are shown as a function of the number of hidden variables. For the clustering, we used 60,000 samples (solid lines) and 20,000 samples (dashed lines). (C) Mean images of each cluster were obtained from our inference method with 60 hidden variables. Difference colors are used just to distinguish different clusters.
}
\end{figure*}

Another potential application of our method, interpreting hidden states as labels, is for unsupervised classification. 
We demonstrate this idea with the MNIST data of handwritten digits~\cite{Lecun1998}.
The data has 60,000 digit samples of 28$\times$28 pixel gray-scale (between 0 and 255) images obtained from 500 different individuals.
Some of sample digits are shown in Fig.~\ref{fig_mnist}A.
Our goal is to classify the 60,000 images into distinct clusters where each cluster represents different digits as well as different writing styles without using true labels.
We formulated the classification problem as follows.
Different digits and writing style combinations are encoded in hidden variable states $\bm{\sigma}^h$.
Then, one realization of $\bm{\sigma}^h(t)$ generates a digit image $\bm{\sigma}^v(t)$, where $t$ is now being used to index the MNIST images.
The feature has $N_h$ degrees of freedom with $\sigma^h_J(t), J=1,\cdots, N_h$.
In particular, for simplicity, we adopted one-hot encoding by assigning only one nonzero element $\sigma^h_J(t)=1$ among $N_h$ elements of $\bm{\sigma}^h(t)$.
Then, the generated image has binary values of $\sigma^v_i(t)=1$ (gray $>1$) or $\sigma^v_i(t)=-1$ (otherwise) for the $i$th pixel, which is determined by the conditional probability,
\begin{equation}
\label{eq:MNIST_Prob}
P(\sigma^v_i(t) = \pm 1 | \bm{\sigma}^{h}(t)) = \frac{\exp ( \pm H_i(t))}{\exp (H_i(t)) + \exp ( - H_i(t))}
\end{equation}
where $H_i(t) \equiv \sum_J W_{iJ}  \sigma_J^{h}(t)$ represents a local field acting on the $i$th pixel.
Here, we ignored observed pixels $i$ if more than 95\% samples had the same value.
The threshold 95\% showed similar results as a more restrictive threshold of 99\%. 
Thus, for this setup, our reconstruction method considers only hidden-to-observed interactions.
Briefly summarizing the inference procedure, we (i) assign a random binary vector $\bm{\sigma}^h(t)$, in which only one element has nonzero value ($\sigma_J^h(t)=1$); (ii) apply  FEM  to reconstruct the interaction strength $W_{iJ}$ from hidden label $J$ to observed pixel $i$; (iii) update the hidden states by assigning $\sigma_J^h(t)=1$ for the label $J$ that makes the likelihood of the observed pixels of sample $t$ the highest 
and $\sigma_J^h(t)=0$ for the other $N_h - 1$ elements; (iv) repeat steps (ii) and (iii) until the discrepancy $D_v$ between $\sigma^v_i(t)$ and $\langle \sigma^v_i(t) \rangle_{\text{model}}$ saturates.
Then, the one-hot hidden states $\bm{\sigma}^h$ represent distinct classes of MNIST images $\bm{\sigma}^v$.

We examined various possible numbers (10 to 100) of labels 
by controlling the number $N_h$ of hidden variables.
As the hidden degrees of freedom $N_h$ increased, the model  generated  images of $\sigma^v_i(t)$ closer to the originals. In other words, the discrepancy $D_v$ kept decreasing as $N_h$ increased (Fig.~\ref{fig_mnist}B).
However, once the model complexity was penalized with the overuse of the hidden degrees of freedom, an optimal degrees of freedom $N_h^*$ was determined with a minimum overall discrepancy $D.$
The estimate $N_h^*\approx 60$ means that the 60,000 MNIST images can be optimally clustered into about 60 classes of digits and writing styles.
The mean images $1/N_c \sum_{t \in c} \sigma^v_i(t)$ corresponding to the 60 labels are shown in Fig.~\ref{fig_mnist}C. Here $N_c$ is the number of samples corresponding to label $c.$
It is of particular interest that each digit was divided into approximately six classes, suggesting that, in the MNIST dataset, about six different writing styles exist  for every digit. To confirm the robustness of this result, we repeated the analysis with only 20,000 of the MNIST images, and obtained a similar conclusion (dashed lines in Fig.~\ref{fig_mnist}B).


\section{Summary}

Given partial observations of systems, complete network reconstruction is a longstanding problem in inference.
In this paper, we propose a new iterative approach based on free energy minimization (FEM) and expectation maximization.
We demonstrated on simulated systems that our method can accurately estimate the actual number of hidden variables from partial observations. Furthermore, network reconstruction was successful in recovering not only observed-to-observed interactions but also those involving hidden variables (hidden-to-observed, observed-to-hidden, and hidden-to-hidden). Hidden-to-hidden interactions are challenging to reconstruct with mean-field methods~\cite{Roudi2013pre, Hertz2014}.
We applied this method to reconstruct a real neuronal network and a stock market network with the inclusion of possible hidden variables. The reconstructed networks were then validated by reproducing real neuronal activities and by a profitable trade simulation, respectively.
Finally, as another potential application to unsupervised pattern classification, we found hidden labels in hand-written digit data.

%
FEM is more effective for network reconstruction than maximum likelihood estimation (MLE), because it separates the cost function evaluation from the independent multiplicative parameter update. This has two major benefits that are crucial for the application to hidden variable problems to succeed. The first is that the cost function can be used as a stopping criterion to avoid overfitting for small sample sizes, important when considering large numbers of possible hidden variables. The second is that the multiplicative update is computationally much more efficient (approximately 100 times faster than usual MLE-based network reconstruction methods), also critical for determining the configurations of hidden variables. 
Since the algorithm reconstructs interactions strengths $W_{ij}$ from the $j$th node to the $i$th node independently for the $i$th node, the network reconstruction can be easily parallelized, and therefore scaled to large system sizes.

%

\section*{Acknowledgment}
This work was supported by Intramural Research Program of the National Institutes of Health, NIDDK (D.-T.H.,V.P.), and by Basic Science Research Program through the National Research Foundation of Korea (NRF) funded by the Ministry of Education (2016R1D1A1B03932264) (J.J.).

\bibliography{hidden}

\end{document}